\documentclass[twocolumn,aps,amsmath,amssymb,superscriptaddress,prl]{revtex4-1}
\usepackage{graphicx,bm,bbm,amsmath,amssymb,natbib,color}
\usepackage{hyperref}
\hypersetup{hidelinks=true,
colorlinks=true,
linkcolor=blue,
urlcolor=blue,
citecolor=blue}

\begin{document}

\title{Electron-Tunneling-Assisted Non-Abelian Braiding of Rotating Majorana Bound States}


\author{Sunghun Park} \email{sunghun.park@uam.es}
\affiliation{Departamento de F\'{\i}sica Te\'orica de la Materia Condensada, Condensed Matter Physics Center (IFIMAC) and Instituto Nicol\'as Cabrera, Universidad Aut\'onoma de Madrid, 28049 Madrid, Spain}
\author{H.-S. Sim}  \email{hssim@kaist.ac.kr}
\affiliation{Department of Physics, Korea Advanced Institute of Science and Technology,  Daejeon 34141, Korea}
\author{Patrik Recher} \email{p.recher@tu-braunschweig.de}
\affiliation{Institute for Mathematical Physics, TU Braunschweig, D-38106 Braunschweig, Germany}
\affiliation{Laboratory for Emerging Nanometrology Braunschweig, D-38106 Braunschweig, Germany}

\date{\today}


\begin{abstract} 
It has been argued that 
fluctuations of fermion parity are harmful for the demonstration of non-Abelian anyonic statistics. Here, we demonstrate a
striking exception in which such fluctuations are actively used.
We present a theory of coherent electron transport from a tunneling tip into a Corbino geometry Josephson junction where four Majorana bound states (MBSs) rotate. 
While the MBSs rotate, electron tunneling happens from the tip to one of the MBSs thereby changing the fermion parity of the MBSs.
The tunneling events in combination with the rotation allow us to identify a novel braiding operator that does not commute with the braiding cycles in the absence of tunneling, revealing the non-Abelian nature of MBSs. The time-averaged tunneling current exhibits resonances as a function of the tip voltage with a period that is a direct consequence of the interference between the non-commuting braiding operations. Our work opens up a possibility for utilizing parity non-conserving processes to control non-Abelian states. 
\end{abstract}

\maketitle

{\it Introduction.---}
A braiding operation reveals the quantum statistics of identical particles~\cite{Nayak2008,Wilczek2009,Stern2010}. Majorana zero-energy states bound to certain defects (e.g. vortices or edges) in topological superconductors are quasiparticles obeying non-Abelian statistics~\cite{Hasan2010,Alicea2012,Beenakker2013,DasSarma2015,Aguado2017}. In an isolated system with $2N$ decoupled Majorana states, there is a $2^{N}$-fold degenerate ground state manifold $\{|\Psi\rangle\}$, and adiabatically moving one Majorana state around another acts as a unitary matrix on the manifold.
Such unitary matrices of different braiding operations, $A$ and $B$, are in general non-commutative, so that the order of operations matter, 
\begin{equation}
AB|\Psi\rangle \neq BA|\Psi\rangle ~~\text{or}~~
(AB-BA)|\Psi\rangle \neq 0. \label{Nonabelian}
\end{equation} 
Non-Abelian braiding is one of the hallmarks of topological quantum phases associated with non-Abelian statistics appearing in many contexts~\cite{Stern2010,Zhu2011,Thomas2016} and also represents the basic resource for executing topologically protected gates for quantum computing~\cite{Nayak2008,Aasen2016}. 

\begin{figure}
\centering
\includegraphics[width=0.98\columnwidth]{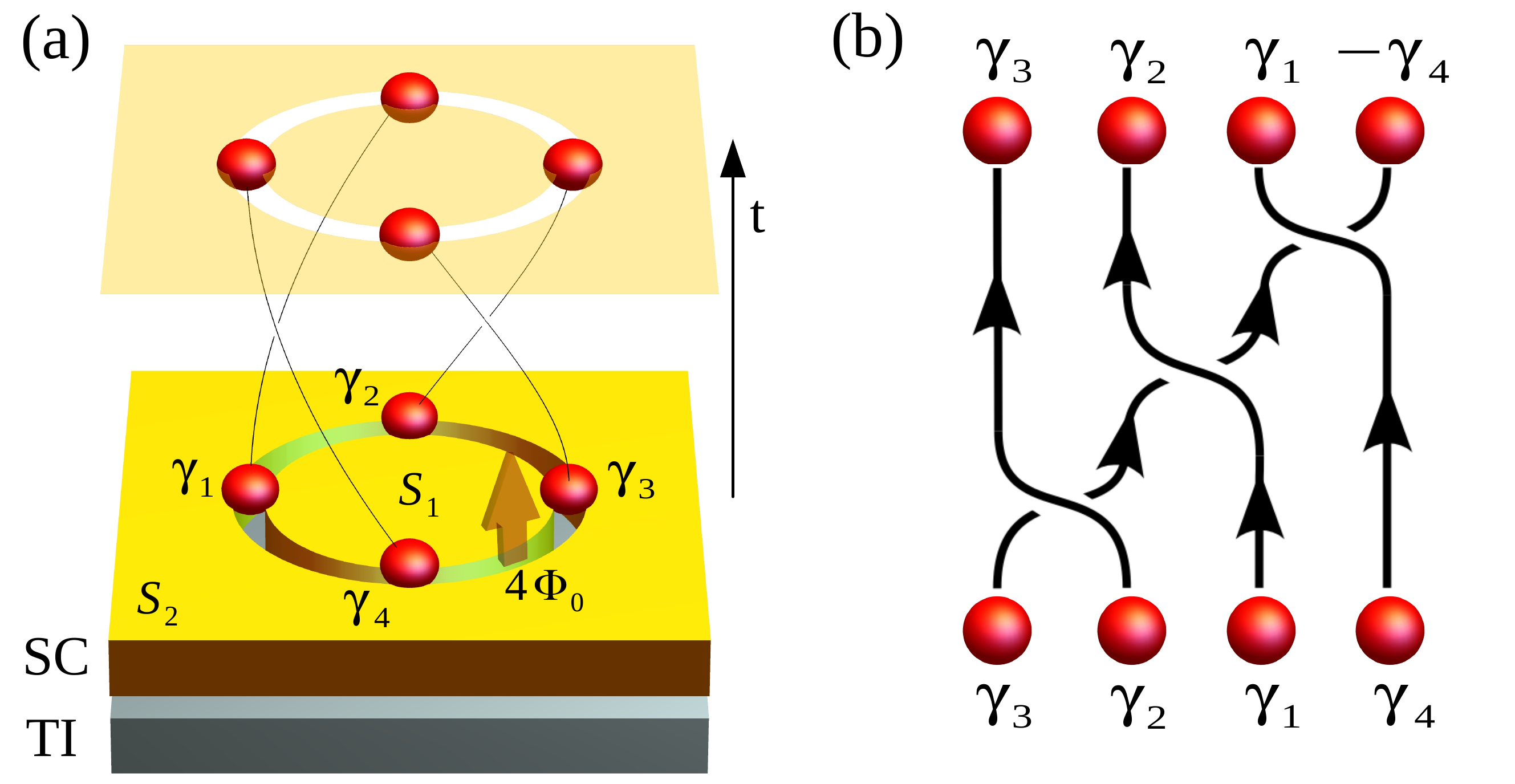}
\caption{(a) Schematic of a Corbino geometry Josephson junction formed by thin-film superconductors ($S_1$ and $S_2$) deposited on the surface of a topological insulator (TI). In the presence of four flux quanta 4$\Phi_0$, four MBSs $\gamma_j$ (red balls) appear in the junction. Majorana positions can move along the junction when applying a small voltage across the junction, allowing us to perform an adiabatic rotation. (b) Braiding depicted as worldlines of the four MBSs corresponding to the $\pi/2$-rotation shown in (a).    
}
\label{Fig1:setup}
\end{figure}

The essence of the present work is to provide transport signatures of Majorana bound states (MBSs) induced by the non-commutativity shown in Eq.~\eqref{Nonabelian}. 
The envisioned system is
a Corbino geometry topological Josephson junction (JJ), formed by two $s$-wave superconductors on a topological insulator (TI) surface [see Fig.~\hyperref[Fig1:setup]{1(a)}].
Four vortices, each hosting a MBS, rotate along the junction, and 
the time-dependent tunneling conductance between the junction and a metallic tip is measured~\cite{Park2015}.  
A ground state of the system evolves in the fourfold degenerate ground state manifold, governed by the rotation and the coherent electron tunneling processes. The evolution can be cast into two braiding operators (corresponding to $A$ and $B$ in Eq.~\eqref{Nonabelian}) which do not commute:  one is a parity-conserving rotation and the other is a tunneling-assisted braiding.
In the low bias voltage regime, the time-averaged conductance exhibits unusual peak positions, which we interpret as a direct signature of non-commutativity of the two braiding operators. 

Tremendous amounts of proposals and experiments lead to great achievements in the realization~\cite{Fu2008,Lutchyn2010,Oreg2010,Alicea2010,Choy2011,Perge2013,Hell2017}, manipulation~\cite{Alicea2011,Flensberg2011,vanHeck2012,Grosfeld2011,Mi2013,Li2016-1} and detection~\cite{Mourik2012,Das2012,Deng2012,Lee2014,Xu2015,Deng2016,Pawlak2016,Feldman2017,Gul2018,Rokhinson2012,Wiedenmann2016,Deacon2017} of MBSs in superconducting hybrid structures. In particular, a recent experiment exploiting a quantum anomalous Hall insulator-superconductor structure~\cite{He2017} 
boosts interest in searches for transport signatures of non-Abelian braiding~\cite{Lian2017, Beenakker2018}. Based on such hybrid structures, the authors of Refs.~\cite{Lian2017,Beenakker2018} theoretically investigated transport properties of Mach-Zehnder-like interferometers of chiral Majorana modes. 
The overlap or fusion of two paths of Majorana modes whose relative dynamics is determined by braiding with the other Majoranas signals a unitary evolution (which is not a phase factor) of Majorana modes. 

Different to these recent studies in Refs.~\cite{Lian2017,Beenakker2018}, we demonstrate interference involving four rotating MBSs whose braiding operations are assisted by tunneling of electrons into or out of the MBSs and thus in which the fermion parity formed by the MBSs is not conserved. Such tunneling-assisted braiding has been to the best of our knowledge not considered before, on the contrary, electron tunneling was seen detrimental for topological quantum processing \cite{Budich2012,Leijnse2012,Woerkom2015}. We will show that, in our scheme, electron tunneling probes non-Abelian statistics via the tunneling conductance.
Our scheme does not require control of fusions of Majorana states.

{\it Theoretical model.---} We consider a Corbino JJ deposited on the surface ($x$-$y$ plane) of a three dimensional TI [Fig.~\hyperref[Fig1:setup]{1(a)}]. The circular shaped junction with a radius $R$ is formed by thin films of inner ($S_1$) and outer ($S_2$) {\it s}-wave superconductors and contains four magnetic flux quanta, $4 \Phi_0$ with $\Phi_0 = h/(2e)$, inducing a  phase difference across the junction (see Eq.~\eqref{SCgap}). The Bogoliubov-de Gennes (BdG) Hamiltonian for the TI surface proximity coupled to the Corbino JJ is given by~\cite{app1}
\begin{align}
H_{C} &= \frac{1}{2} \int d^2 r ~\Phi^{\dagger}({\mathbf r}) \mathcal{H}_C\Phi({\mathbf r}), \label{Hamiltonian} \\
\mathcal{H}_C &= 
\left(
 \begin{array}{ccc}
  \mathcal{H}_0 - \mu & \,\, \Delta({\mathbf r})  \\
  \Delta^{*}({\mathbf r}) & \,\, \mu - \mathcal{H}_0
 \end{array}
\right),
\end{align}
and $\Phi({\mathbf r}) = (\Phi_{\uparrow},\Phi_{\downarrow}, \Phi^{\dagger}_{\downarrow},-\Phi^{\dagger}_{\uparrow})^{T}$ is the Nambu spinor and $\mathcal{H}_0 = v_F (\sigma_x p_x + \sigma_y p_y)$ with Pauli spin matrices $\sigma_{x,y}$ describes the surface states and $\mu$ is the chemical potential. The proximity-induced superconducting gap $\Delta({\mathbf r})$ is 
\begin{align}
\Delta({\mathbf r})= 
\left\{ 
 \begin{array}{cc}
  \Delta_0 e^{i \phi_1} & 0 \leq r < R,\\
  \Delta_0 e^{-i 4 \theta + i \phi_2} &  r > R,
 \end{array}
\right.\label{SCgap}
\end{align}
where $\phi_1$ and $\phi_2$ are spatially uniform phases in each superconducting region, and the polar-angle-dependent phase $-4 \theta$ at $r>R$ is due to the presence of the four flux quanta~\cite{Clem2010}. By solving the BdG equation $\mathcal{H}_C \Psi({\mathbf r})= E \Psi({\mathbf r})$, we find four Majorana wave functions $\Psi_{\text{M}j}({\mathbf r})$ with $j \in \{1, 2, 3, 4 \}$, at zero energy $E=0$. They are localized at ($r,\theta$) = ($R, \theta_j$) where $\theta_j = (3\pi - 2\pi j)/4 - (\phi_1-\phi_2)/4$, at which the local phase difference across the junction is $\pi$. Detailed calculations of the Majorana wave functions for $\mu=0$ are given in Supplemental material~\cite{Suppl}.   

If we change $\phi_1 - \phi_2$ by $2 \pi$, the four MBSs rotate by $\pi/2$ in a clockwise direction maintaining their relative distances, as plotted in Fig.~\hyperref[Fig1:setup]{1(a)}, leading to a transformation $\gamma_j \rightarrow U_c \gamma_j U^{\dagger}_c$,
\begin{eqnarray}
\begin{split}
 \gamma_{1} \rightarrow -s\gamma_{2},\hspace{12pt}
 \gamma_{2} \rightarrow -s\gamma_{3}, \\
 \gamma_{3} \rightarrow  s\gamma_{4}, \hspace{12pt}
 \gamma_{4} \rightarrow -s\gamma_{1}, 
\end{split}\label{C_opt}
\end{eqnarray}
where $\gamma_{j}= \int d^2 r ~\Psi^{\dagger}_{\text{M} j} ({\mathbf r}) \Phi({\mathbf r})$. $s = 1 (-1)$ corresponds to the change of $\phi_1 (\phi_2)$ by $2\pi (-2\pi)$. Graphical representation of the transformation is given in Fig.~\hyperref[Fig1:setup]{1(b)} for the $s=-1$ case. A rotation operator $U_c$ for the transformation can be constructed as a product of three pairwise braidings $U_c = U_{41} U_{12} U_{23}$ where $U_{ij}$ is the braiding exchange operator of $\gamma_i$ and $\gamma_j$ given by $U_{ij} = \text{exp}\left(s \pi \gamma_i \gamma_j/4\right)$~\cite{Ivanov2001}.

The adiabatic rotation can be achieved if a dc-bias voltage $V_J$ across the junction is much smaller than the excitation energy of the junction. For a finite $V_J$, $\phi_1 - \phi_2$ varies in time $t$ as $\phi_1 - \phi_2 = \phi_0 + 2 e V_J t/\hbar$ where $\phi_0$ is a spontaneously chosen constant. The states $\Psi_{\text{M}j}({\mathbf r},\phi_1(t), \phi_2(t))$ then become instantaneous eigenstates of $\mathcal{H}_{C}[\phi_1(t), \phi_2(t)]$ at zero energy, and $U_c$ can be considered as the time evolution operator of the MBSs from $t$ to $t+T_J$, where $T_J = \frac{\pi \hbar}{e V_J}$ is the time needed for the $\pi/2$-rotation.

\begin{figure}
\includegraphics[width=0.98\columnwidth]{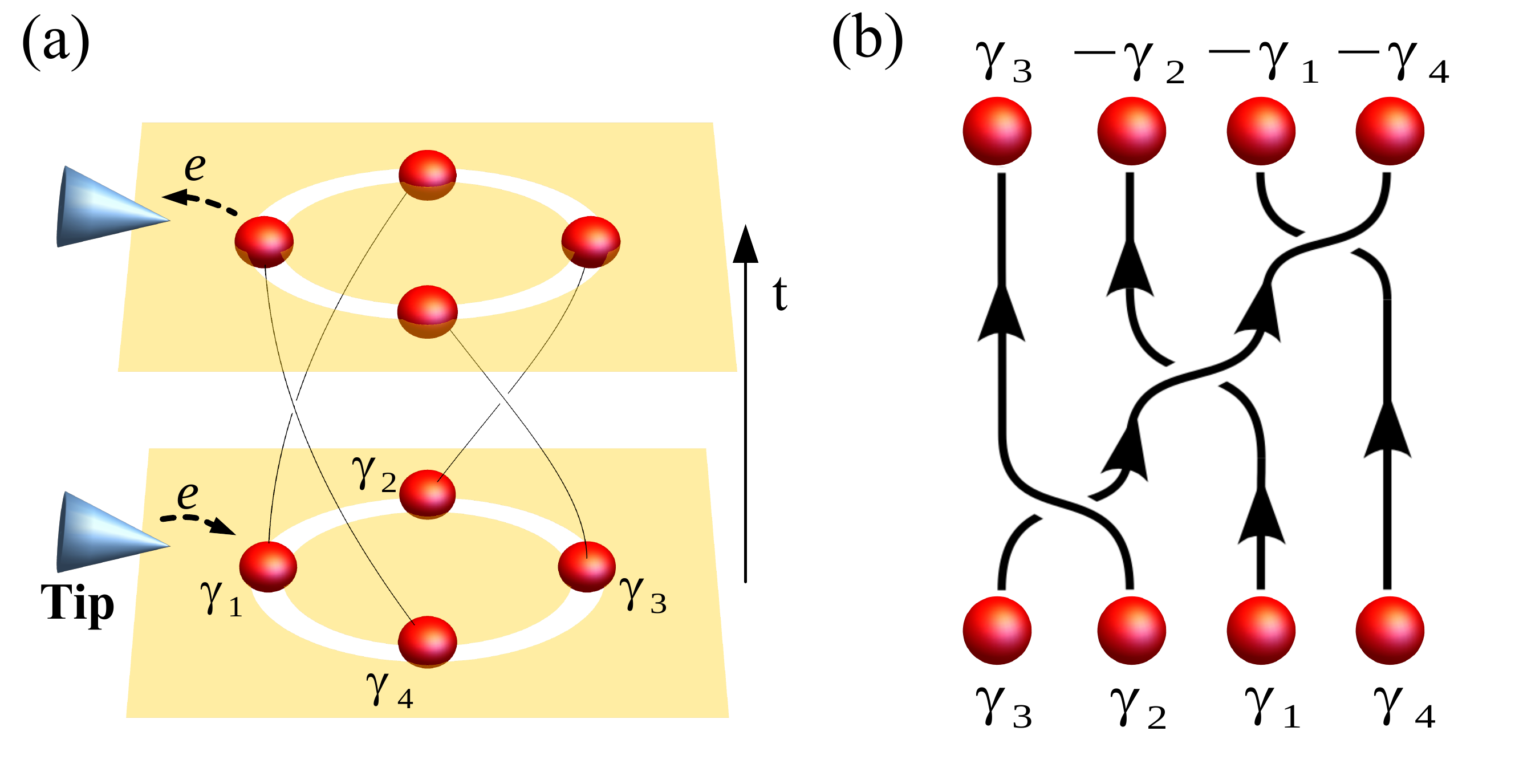}
\caption{(a) Time-dependent electron tunneling between the rotating MBSs and a metal tip for detecting non-Abelian statistics. (b) Tunneling-assisted braiding created by the composition of the $\pi/2$-rotation shown in (a) and electron tunneling. The tunneling effect reverses the exchange direction of a Majorana pair involving $\gamma_1$. A signature of the interference processes involving the non-Abelian braiding operations --- the tunneling-assisted braiding and the braiding shown in Fig.~\hyperref[Fig1:setup]{1(b)} --- is probed by the time-averaged tunneling current. 
}
\label{Fig2:tunneling}
\end{figure}

{\it Tunneling-assisted Majorana braiding.---} 
To explore the effect of electron tunneling, we connect a metal tip to the Corbino JJ, as depicted in Fig.~\hyperref[Fig2:tunneling]{2(a)}. The tip is located such that an electron can tunnel onto or off the Corbino JJ through $\gamma_1(t_0)$ at $t=t_0$, and we assume that the tunnel coupling is switched on at $t=t_0$. A phase coherent time-dependent tunneling event between the tip and adiabatically rotating Majorana states can occur at discrete times $t_q = t_0 + q T_J$, where $q=0, 1, 2,...$. Creation or annihilation of an electron via a Majorana state at $t=t_q$ is described by $\gamma_1(t_0)|\Psi_g(t_q)\rangle$ where $|\Psi_g(t_q)\rangle=U_c^{q}|\Psi_g(t_0)\rangle$ is the time-evolved initial state (being part of the ground-state manifold) of the MBSs from $t_0$ to $t_q$. Note that our proposal does not depend on the initial configuration of the ground state and other choices of Majorana states coupled to the tip at $t=t_0$. Hereafter, we will denote $\gamma_1(t_0)$ by $\gamma_1$. 

The time evolution of a Majorana state from $t=t_{q'}$ to $t_q$ at which tunneling events occur is described by the Majorana Green's function
\begin{align}
M(t_q,t_{q'})= -i~ \text{Tr}\left[ \rho_0 \, \hat{\gamma}_1(t_q) \hat{\gamma}_1(t_{q'})\right], \label{MGF}
\end{align}
where $\hat{\gamma}_1(t_q)= \left(U^{\dagger}_c\right)^{q} \gamma_1 U_c^{q}$ and $\rho_0$ is a density matrix of the Majorana state at $t=t_0$. 
For a more comprehensive description of the tunneling effect, we introduce a tunneling-assisted braiding operator, 
\begin{align}
\bar{U}_c = \gamma_1 U_c \gamma_1, \label{Ubar_opt}
\end{align}
consisting of three events: changing fermion-occupation-number parity due to the tunneling at $t=t_q$, followed by an evolution for a time $T_J$ with $U_c$, and then changing the parity again at $t=t_q + T_J$. The transformation governed by $\bar{U}_c$ is drawn in Fig.~\hyperref[Fig2:tunneling]{2(b)}; comparing the cases without and with the tunneling in
Figs.~\hyperref[Fig1:setup]{1(b)} and \hyperref[Fig2:tunneling]{2(b)}, respectively, notice that
the tunneling effectively reverses the direction of the pairwise braiding when a braiding involves $\gamma_1$. Therefore, ${\bar U}_c=U_{14}U_{21}U_{23}$ can be considered -- besides $U_c$ -- as another genuine braiding operator.
$M(t_q,t_{q'})$ then can be presented as 
\begin{align}
M(t_q,t_{q'})= -i~ \text{Tr}\left[ \rho'_0 \, \left(\bar{U}_c\right)^{n}\left(U^{\dagger}_c\right)^{n}\right], \label{MGF2}
\end{align}
where we used the cyclic property of the trace. $\rho'_0 = \left(U_c\right)^{q}\rho_0 \left(U^{\dagger}_c\right)^{q}$ and $n=q-q'$. We find that $U^{\dagger}_c$ and $\bar{U}_c$ do not commute, $\left[\bar{U}_c, U^{\dagger}_c \right]\neq 0$. As a consequence, $M(t_q,t_{q\prime})$ is not just a sum of phase factors but involves non-trivial state changes in the ground-state manifold. We show below that the non-commuting braidings result in observable interference signatures free of the necessity of physically fusing MBSs. 

{\it Transport signatures.---} 
To obtain the tunneling current between the tip and the JJ in the weak coupling limit, 
we extend the formalism of Ref.~\cite{Park2015} to four MBSs. The Hamiltonian of the tip is $H_N = \sum_{k\sigma} \varepsilon_k c^{\dagger}_{k\sigma} c_{k\sigma}$ where $c_{k\sigma}$ is the electron annihilation operator in the tip with momentum $k$ and spin $\sigma$. Since we are interested in the low-energy sector of the junction, tunneling between the tip and the MBSs is the only relevant process. Around $t=t_q$ where the coupling strength to $\gamma_1$ is maximal, we assume that the coupling increases and decreases exponentially as $\gamma_1$ approaches to and leaves from the tip, respectively, while its phase does not change significantly. Moreover, since the Majorana states are spin polarized, and couple only to electrons of the tip with their spin parallel to that of the Majorana states; electrons with opposite spin are reflected at the junction between the tip and the Corbino JJ and do not contribute to the tunneling current. Then the tunneling Hamiltonian becomes 
\begin{align}
 H_T(t) &= \sum_{k,q}  e^{-\lambda |t-t_q|} V_{1k} \, c^{\dagger}_k \gamma_1 + \text{H.c.},\label{TunnelingH}
\end{align} 
where $\lambda^{-1}$ is the tunneling duration and $V_{1k}$ is the coupling between the tip and $\gamma_1$. Here we have assumed $\lambda^{-1} \ll T_J$, implying that only nearest-neighbor coupling between the tip and the MBSs is taken into account. 

Using the current expression $ I (t) = - e\, d N_T/dt$ with the tip number operator $N_T = \sum_{k} c^{\dagger}_{k} c_{k} $ and lowest order perturbation theory in $H_T(t)$, the differential conductance of the time-averaged current measured after many rotation cycles 
of MBSs has the form,  
\begin{figure}
\includegraphics[width=0.95\columnwidth]{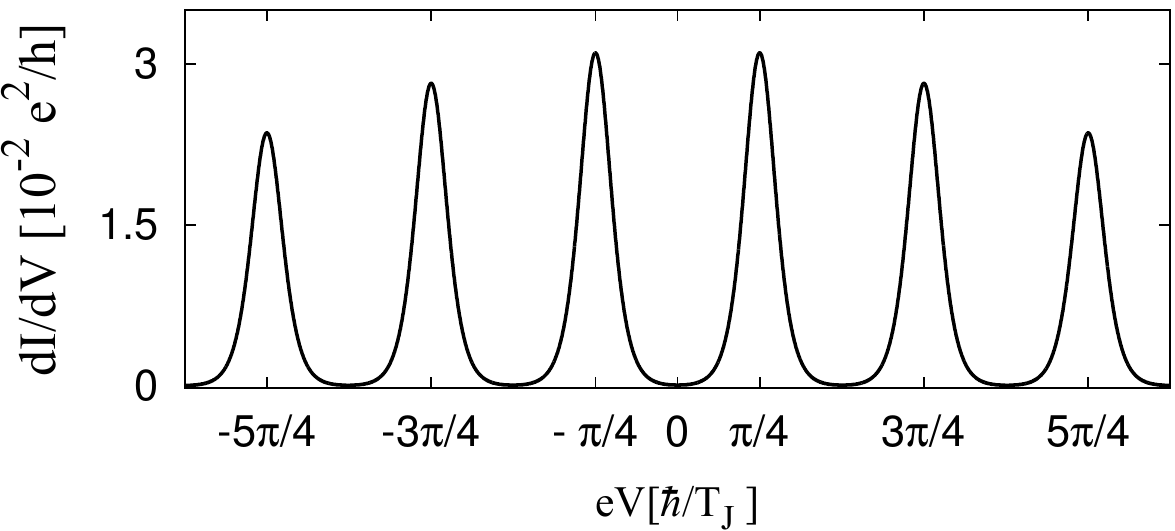}
\caption{Plot of the time averaged differential conductance given in Eq.~\eqref{dIdV} 
with parameters $\hbar T_{J}^{-1} = 0.1 \text{meV} = 10^{-1} \hbar \lambda = 10 k_B T = 10 \Gamma$. The conductance peak spacing $h/(4 T_J)$ is a consequence of the non-Abelian state evolution within the degenerate ground-state manifold.  
}
\label{Fig3:dIdV}
\end{figure}
\begin{align}
\frac{d\bar{I}}{dV}
&=\frac{e}{h} \int^{\infty}_{-\infty}d\varepsilon \, 
T(\varepsilon) \left[S(\varepsilon) + S(-\varepsilon)\right] \frac{d n_F(\varepsilon - eV)}{dV}, \label{dIdV_form}  
\end{align} 
where $n_F$ is the Fermi-Dirac distribution and $eV$ is the bias voltage. The tunneling probability $T(\varepsilon)$ and the interference term $S(\varepsilon)$ are given by
\begin{align}
T(\varepsilon)&= \frac{2 \Gamma T_J}{\hbar} \left(\frac{2 \lambda T_J}{\lambda^2 T^2_J + \tilde{\varepsilon}^2}\right)^2, \label{Tform}\\
S(\varepsilon)&= \text{Re}\left\{\frac{1}{2}+i \sum^{Q}_{n=1}e^{i n \tilde{\varepsilon}}
 M(t_Q,t_{Q-n}) \right\}. \label{Sform}
\end{align}
Here $\tilde{\varepsilon}=\varepsilon/(\hbar T^{-1}_J)$ and the integer $Q \gg 1$, which will go to infinity later. $\Gamma = 2\pi \rho |V_{1k}|^2$ where $\rho$ is the tip density of states. We assumed a wide-band approximation where $\rho$ and $V_{1k}$ are energy independent and 
we neglected the contributions proportional to $e^{-\lambda T_J/2}$; note that these small contributions do not change the positions of conductance peaks. The details for the calculation of $\bar{I}$ are given in \cite{Suppl}. In the limit $Q \rightarrow \infty$, we obtain
\begin{align}
\frac{d\bar{I}}{dV}= \frac{e^2}{h}\frac{\pi \hbar}{8 T_J k_B T} 
\sum_l T(\varepsilon_l)~\text{sech}^2\left(\frac{eV - \varepsilon_l}{2 k_B T}\right),\label{dIdV}
\end{align}
which shows peaks at $\varepsilon_l=\frac{\hbar}{4T_J}(2\pi l-\alpha)$ where $l$ is an integer and $\alpha=\pi$ arising from a $2\pi$-rotation of the four MBSs.  
This perturbative calculation is valid for $T(\varepsilon_0) \hbar/(8 T_J) \ll k_B T \ll E_g$ 
where $E_g$ is the excitation energy of the junction. 

The $d\bar{I}/dV$ in Eq.~\eqref{dIdV} is plotted in Fig.~\ref{Fig3:dIdV} for realistic parameters. It shows peaks at $eV = \varepsilon_l$. This is our main result. The peak positions are determined by $T_J$ and $\alpha$, but are independent of system details such as the initial Majorana state at $t=t_0$ and the tunneling strength $\Gamma$. Note that the periodicity $T_J$ of the system Hamiltonian in Eqs.~\eqref{Hamiltonian} and \eqref{TunnelingH} does not coincide with the periodicity of the ground state $4 T_J$ from the fact that $U_c^{4} ={\bar U}_c^{4} = \mathbbm{1}$. It is a consequence of the non-trivial state evolution within the ground-state manifold of 4MBSs requiring a matrix structure. As shown below, the $4 T_J$-periodicity and the non-commutativity between $U_c$ and $\bar{U}_c$ result in peaks in $d{\bar I}/dV$ separated by $h/(4T_J)$ and not by $h/T_J$ associated with the frequency of appearance of MBSs beneath the tip. The results are the same for the case of an anticlockwise rotation of four MBSs.

{\it Non-Abelian statistics.---} 
In order to clearly unveil such a link between the interference effect and the non-Abelian matrix structure, we analyse the term $S(\varepsilon)$ in the occupation number basis $\{|n_1 n_2\rangle\}$, where $n_1, n_2 = 0,1$ are occupation numbers for fermionic operators $f_1 =\frac{\gamma_1 + i \gamma_2}{2}$ and $f_2 =\frac{\gamma_3 + i \gamma_4}{2}$, see \cite{Suppl} for more details on the occupation number representation. As Eq.~\eqref{dIdV} 
is independent of the initial condition, the specific form of the initial density matrix ($\rho_0$ or $\rho'_0$) is unimportant. Substituting Eq.~\eqref{MGF2} into Eq.~\eqref{Sform} leads to $S(\varepsilon) = \text{Re}\{ \text{Tr} [\rho'_0 \hat{S}(\varepsilon)]\}$ where 
\begin{align}
\hat{S}(\varepsilon) = \frac{1}{2} + \sum^{Q}_{n=1} 
e^{i n \tilde{\varepsilon}}  \left(\bar{U}_c\right)^{n} \left(U^{\dagger}_c\right)^{n}. \label{Sopt}
\end{align}
Note that the operations $U_c^\dagger$ and $\bar{U}_c$ do not commute, and thus the sum {\it cannot} be treated as a simple geometric series: $\sum^{\infty}_{n=1} e^{i n \tilde{\varepsilon}} 
\left(\bar{U}_c\right)^{n} \left(U^{\dagger}_c\right)^{n} \nrightarrow 
\sum^{\infty}_{n=1} e^{i n \tilde{\varepsilon}} e^{i n \varphi}$. The operator $\hat{S}(\varepsilon)$ comes from the overlap between the following two processes of temporal length $Q T_J$: In process I,  an electron tunnels from the tip to $\gamma_1$ at $t_0 + (Q-n) T_J$, and in process II, the tunneling happens at $t_0 + QT_J$. Here $e^{i n \tilde{\epsilon}}$ is the dynamical phase factor gained for the time interval $n T_J$. The interference between terms of different $n$ determines the peak positions of the conductance. 

Let us assume that an even parity state, a mixture of $|00\rangle$ and $|11\rangle$, is prepared at $t=t_0$; the case of an odd parity state is obtained in a similar way. In the limit $Q\rightarrow \infty$, Eq.~\eqref{Sopt} for an even parity is given by 
\begin{align}
\hat{S}(\varepsilon)\big|_{\text{even}} 
&= \frac{1}{2}+\big(-si\tau_z e^{i \tilde{\varepsilon}} 
-i\tau_y e^{i 2 \tilde{\varepsilon}}+s i \tau_x e^{i 3 \tilde{\varepsilon}} \nonumber\\
&\hspace{35pt}-e^{i 4 \tilde{\varepsilon}}
\big)\times \sum^{\infty}_{m=0} e^{i m(4 \tilde{\varepsilon}+\pi)},\label{Sopt2}
\end{align}
where $\tau_{x,y,z}$ are Pauli matrices acting in the space of the even parity states, $|00\rangle$ and $|11\rangle$. In the second line, the summation is classified into four categories in each of which the Pauli matrix (including the identity matrix) is factored out, manifesting the interference with period $4 T_J$. Using Eqs.~(\ref{Sopt}) and (\ref{Sopt2}) yields $S(\varepsilon)+S(-\varepsilon)=\sum_m\exp(im(4{\tilde \varepsilon}+\pi))\sim\sum_l\delta(4{\tilde \varepsilon}+\pi(2l+1))$, where $m,l$ are integers. Together with Eq.~(\ref{dIdV_form}) we obtain our final result Eq.~(\ref{dIdV}). We note that the period of $4 T_J$ cannot be obtained by corresponding braiding operators that would commute, see \cite{Suppl}. We also note that this non-Abelian interference effect cannot be envisaged in a system with two MBSs where non-commuting braiding operations do not occur \cite{Park2015}.

We remark that the suggested test of non-Abelian braiding statistics needs only a local measurement of MBSs that are at zero energy so that the way we fuse the 4 MBSs into the two fermions $f_1$ and $f_2$ is actually arbitrary. The period $4 T_J$ also does not depend on a specific initial state  (if the time-average is performed after times $t \gg T_J$) but is only a consequence of the non-commuting matrix structure of $U_c$ and ${\bar U}_c$. The extracted information of the state changes is due to interference that is generated because the MBSs rotate in the Corbino geometry JJ. This is  fundamentally different compared to other braiding schemes which use the selective switching on and off of couplings between the Majorana bound states and the read-out of the non-Abelian state changes is done without physically moving the MBSs~\cite{vanHeck2012,Bonderson2013}. In our scheme the rotation induces a {\it dynamical coupling} between the MBSs as we discuss in detail in the Supplemental material employing the Floquet picture. There we consider also the zero temperature case to all orders in the tunneling from the tip to the MBSs.

{\it Discussion and conclusion.---} 
We have demonstrated that a non-Abelian state evolution can be identified in tunneling conductance measurements between four rotating MBSs in a Corbino geometry topological Josephson junction and a metal tip. Unitary evolutions of the MBSs acting on even and odd parity subspaces, which are separable if the fermion parity is conserved, are intertwined by electron tunneling, inducing parity-conserving and tunneling-assisted braiding operators. Coherent interference between different orders of round trips of Majorana states governed by the parity-conserving and tunneling-assisted braiding operators yields a time-averaged conductance exhibiting peaks with a period of $h/(4T_J)$ as a function of bias voltage between the metal tip and the Josephson junction, whereas the period of the Hamiltonian is $T_J$. This constitutes a clear signature of non-Abelian state evolution of four MBSs. 

We explicitly showed that these results have their origin in the non-commutativity of the parity-conserving and tunneling-assisted braiding operators and are therefore independent on the way we fuse the MBSs into fermions which is fundamentally different from other recent proposals that use time-dependent couplings between the MBSs or Coulomb interaction to lift their degeneracies~\cite{vanHeck2012,Bonderson2013,Aasen2016}. Here, an effective coupling between MBSs is induced dynamically by the rotation which only requires a dc-Josephson voltage applied between the two superconductors.

We expect that other kinds of exotic zero modes such as MBSs in time-reversal invariant topological superconductors~\cite{Zhang2013,Keselman2013,Haim2014,Wolms2015,Wolms2016,Li2016-2,Schrade2018} and parafermions~\cite{Fendley2012,Lindner2012,Cheng2012,Clarke2013,Vaezi2013,Barkeshli2014,Jelena2014,Maghrebi2015,Alicea2016} could be analyzed with our time-dependent tunneling scheme to manifest the quantum statistics of the corresponding modes.

The experimental realization may be challenging, but within reach of current experiments. Assuming the proximity-induced superconducting gap $\Delta_0=$1 meV that can be achieved, for example, in thin-films of Nb or NbN~\cite{Lin2013,Du2017}, the excitation energy gap of Josephson vortices of the junction can be estimated by $E_g=\Delta_0\sqrt{4 \xi/R}\sim 0.9$ meV for the radius of the junction $R=5 \xi$~\cite{Park2015,Potter2013}, where $\xi$ is the superconducting coherence length. We require a coherent and adiabatic rotation of the MBSs so that $T_J$ (the time taken for the $\pi/2$ rotation) should satisfy $\hbar/E_g (=0.7~\text{ps}) \ll T_J \ll t_{\text{qp}} (\gtrsim \mu\text{s})$ where $t_{\text{qp}}$ is the quasiparticle poisoning time~\cite{Rainis2012,Higginbotham2015}. At the same time, the temperature should be much smaller than the separation between the conductance peaks $h/(4 T_J)$. MBSs can be spaced unequally apart in the presence of inhomogeneities in the junction. However, they do not affect the rotation time $T_J$ due to the periodicity of the system Hamiltonian and corresponding interference traces on the time scale of $4 T_J$ due to non-Abelian evolution would remain. We believe that the Corbino geometry topological Josephson junction can also be realized in heterostructures of a thin-film topological insulator and a superconductor~\cite{Hao2017} or Pb/Co/Si(111) two-dimensional topological superconductor~\cite{Cren2018}.  

Our findings provide a new way of looking at braiding experiments, by actively using parity switching events by tunneling, instead of avoiding them. This may define a new way to build non-Abelian operations for topological qubits utilizing coherent fluctuations of the fermion parity. Such a change of fermion parity could be  achieved on demand during a definite time using charge pumps based on quantum dots in the single electron regime \cite{Fricke2014} coupled to the setup. Quantum dots could already be coupled to MBSs in experiment \cite{Deng2016}.

\acknowledgments  
We thank A. Levy Yeyati for helpful discussion. S.P. is supported by the Spanish MINECO through the ``Mar\'{\i}a de Maeztu'' Programme for Units of Excellence in R\&D (MDM-2014-0377). 
P.R. acknowledges financial support from the Lower Saxony PhD-programme $"$Contacts in Nanosystems$"$, the Braunschweig International Graduate School of Metrology B-IGSM, the $"$Nieders{\" a}chsisches Vorab$"$ through $"$Quantum- and Nano-Metrology (QUANOMET)$"$ initiative within the project NL-2, and the Deutsche Forschungsgemeinschaft (DFG, German Research Foundation) within the Research Training Group GrK1952/1 $"$Metrology for Complex Nanosystems$"$ and the framework of Germany's Excellence Strategy -- EXC-2123 QuantumFrontiers -- 390837967.
H.-S. S. acknowledges support from the National Research Foundation (Korea NRF) funded by the Korean Government via the SRC Center for Quantum Coherence in Condensed Matter (Grant No. 2016R1A5A1008184).

\clearpage

\setcounter{equation}{0}
\setcounter{figure}{0}
\renewcommand{\theequation}{S\arabic{equation}}
\renewcommand{\thefigure}{S\arabic{figure}}
\renewcommand*{\citenumfont}[1]{S#1}
\renewcommand*{\bibnumfmt}[1]{[S#1]}

\widetext
\begin{center}
\textbf{\large Supplemental material for ``Electron-Tunneling-Assisted Non-Abelian Braiding of Rotating Majorana Bound States''}

\bigskip 

Sunghun Park$^1$, H.-S. Sim$^2$, and Patrik Recher$^{3,4}$ 

$^{\it{1}}$\textit{Departamento de F\'{\i}sica Te\'orica de la Materia Condensada, Condensed Matter Physics Center (IFIMAC) and Instituto Nicol\'as Cabrera, Universidad Aut\'onoma de Madrid, 28049 Madrid, Spain}\\
$^{\it{2}}$\textit{Department of Physics, Korea Advanced Institute of Science and Technology,  Daejeon 34141, Korea}\\
$^{\it{3}}$\textit{Institute for Mathematical Physics, TU Braunschweig, D-38106 Braunschweig, Germany}\\
$^{\it{4}}$\textit{Laboratory for Emerging Nanometrology Braunschweig, D-38106 Braunschweig, Germany}\\
\end{center}

\section*{A. Time-averaged tunneling current} \label{TAcurrent}

The time-dependent tunneling current between a metal tip and a Corbino geometry topological Josephson junction in the weak tunneling limit can be obtained using lowest order perturbation theory. To lowest order in $H_T(t)$, we find the tunneling current  $\langle I(t)\rangle=-e \langle d N_T(t)/dt \rangle$, 
\begin{align}
 \langle I(t)\rangle =&\frac{1}{i \hbar} \int^{t}_{t_0} dt'  \langle [\hat{I}(t), \hat{H}_T(t')] \rangle \nonumber\\
    =&\frac{2e}{\hbar^2} \text{Re} \Bigg\{ \int^{t}_{t_0} dt' \sum_{kqq'} \Gamma_{kqq'}(t,t') 
      [G_{k}(t,t')-\bar{G}_{k}(t,t')] M(t,t') \Bigg\}, \label{PerturbationCurrent}
\end{align}
where $N_T(t)= \displaystyle \sum_{k} c^{\dagger}_{k}(t) c_{k}(t)$ is the metal tip number operator. $\hat{H}_T(t')$ and $\hat{I}(t)$ which are expressed in the interaction picture are given by 
\begin{align}
\hat{H}_T(t') &= \sum_{kq}  e^{-\lambda |t'-t_q|} V_{1k}(t_0) \hat{c}^{\dagger}_k (t') \hat{\gamma}_1(t') + \text{H.c.},\\
\hat{I}(t) &= \frac{e}{\hbar} \sum_{kq} [i  e^{-\lambda |t-t_q|} V_{1k}(t_0) \hat{c}^{\dagger}_k (t) \hat{\gamma}_1(t) +\text{H.c.}].
\end{align}
The tunneling Hamiltonian $\hat{H}_T(t')$ switched on at time $t_0$ is valid in the low energy regime where MBSs are the only relevant states for the tunneling current and for $\lambda^{-1} \ll T_J$. The coupling coefficient $V_{1k}(t_0)$ between the tip and $\hat{\gamma}_1(t_0)$ is 
\begin{align}
V_{1k}(t_0) = \int d^2 r~ t_{k}({\mathbf r}) \Psi_{M1 \downarrow} ({\mathbf r},t_0), 
\end{align}
where $t_{k}({\mathbf r})$ is the tunneling coefficient between the tip and the junction and $\Psi_{M1 \downarrow} ({\mathbf r},t_0)$ is the electron spin-down component of the Majorana wave function $\Psi_{M1} ({\mathbf r})$ in Eq.~\eqref{MBSs}.
In Eq.~\eqref{PerturbationCurrent}, the time-dependent tunneling parameter $\Gamma_{kqq'}(t,t')$ and the tip-electron Green's functions $G_{k}(t,t')$ and $\bar{G}_{k}(t,t')$ are given by  
\begin{align}
\begin{split}
\Gamma_{kqq'}(t,t') &= |V_{1k}(t_0)|^2 e^{-\lambda |t-t_q|} e^{-\lambda |t'-t_{q'}|},\\
G_{k}(t,t') &= -i \langle \hat{c}_{k}(t) \hat{c}^{\dagger}_{k}(t')\rangle 
= -i e^{-i (\varepsilon_k + eV)(t-t')/\hbar} \left[ 1 - n_F (\varepsilon_k)\right],\\
\bar{G}_{k}(t,t') &=-i \langle \hat{c}^{\dagger}_{k}(t) \hat{c}_{k}(t')\rangle = 
-i e^{i (\varepsilon_k + eV)(t-t')/\hbar} n_F (\varepsilon_k),
\end{split}
\end{align}
where $\langle \cdot \rangle$ is the expectation value over a thermal ensemble of initial states 
at $t=t_0$, and $n_F(\varepsilon_k) = 1/[1+e^{\varepsilon_k/(k_B T)}]$ is the Fermi-Dirac distribution at $t=t_0$ 
with the temperature $T$.  Since the tunneling current is exponentially small except for $t=t_q$ and $t'=t_{q'}$ due to the presence of the exponential factor of $\Gamma_{kqq'}(t,t')$, we can approximate the Majorana Green's function, 
\begin{align}
M(t,t')\approx M(t_q,t_{q'}) =-i~ \text{Tr}\left[ \rho_0 \, \hat{\gamma}_1(t_q) \hat{\gamma}_1(t_{q'})\right],
\end{align} 
where $\rho_0$ is a density matrix of the Majorana state at $t=t_0$. If the Josephson junction is in one of the ground states $|\Psi_g(t_0)\rangle$ at $t=t_0$, the density matrix has the form of $\rho_0 = |\Psi_g(t_0)\rangle \langle \Psi_g(t_0)|$ and we get 
\begin{align}
M(t_q,t_{q'}) = -i \langle \Psi_{g}(t_0)|\hat{\gamma}_1(t_q) \hat{\gamma}_1(t_{q'})|\Psi_g(t_0)\rangle. \label{MGF_1}
\end{align}
If $t$ is very far from $t_0$, we can find that the difference between $\langle I(t)\rangle$ and $\langle I(t-T_J)\rangle$ is negligible, 
\begin{align}
\langle I(t)\rangle - \langle I(t-T_J)\rangle \sim \sum_k e^{-i (\varepsilon_k + eV)(t-t_0)} n_F(\varepsilon_k) + \text{c.c.} \sim 0, 
\end{align}
yielding a time-periodic behavior of the tunneling current $\langle I(t)\rangle=\langle I(t-T_J)\rangle$.
Without loss of generality, we assume that $t$ is in the interval $\left[(Q-1/2) T_J, (Q+1/2) T_J\right]$ where $Q$ is a very large integer, $Q \gg 1$. Then the time-averaged tunneling current over an interval $\left[ \tilde{t}-T_J, \tilde{t} \right]$ is 
\begin{align}
\bar{I} = \frac{1}{T_J} \int^{\tilde{t}}_{ \tilde{t} - T_J} dt  \langle I(t)\rangle. 
\end{align}
Let us change the variable in Eq.~\eqref{PerturbationCurrent} from $\varepsilon_k$ to $\varepsilon_k-eV$. After some algebra, we find $\bar{I}$ as 
\begin{align}
\bar{I} &= \frac{e}{h} \int^{\infty}_{-\infty} d\varepsilon~ T(\varepsilon)S(\varepsilon) [n_F (\varepsilon - eV) - n_F (\varepsilon + eV)] \nonumber\\
&=\frac{e}{h} \int^{\infty}_{-\infty} d\varepsilon~ T(\varepsilon)S(\varepsilon) n_F (\varepsilon - eV)+ 
\frac{e}{h} \int^{\infty}_{-\infty} d\varepsilon~ T(\varepsilon)S(\varepsilon) n_F (-\varepsilon - eV)- 
\frac{e}{h} \int^{\infty}_{-\infty} d\varepsilon~ T(\varepsilon)S(\varepsilon) \nonumber\\
&=\frac{e}{h} \int^{\infty}_{-\infty} d\varepsilon~ T(\varepsilon) [S(\varepsilon)+S(-\varepsilon)] n_F (\varepsilon - eV)- 
\frac{e}{h} \int^{\infty}_{-\infty} d\varepsilon~ T(\varepsilon)S(\varepsilon), \label{Suppl:Current}
\end{align}
where $T(\varepsilon)$ and $S(\varepsilon)$ are 
\begin{align}
T(\varepsilon)&= \frac{2 \Gamma T_J}{\hbar} \left(\frac{2 \lambda T_J}{\lambda^2 T^2_J + \tilde{\varepsilon}^2}\right)^2,\\
S(\varepsilon)&= \text{Re}\left\{\frac{1}{2}+i \sum^{Q}_{n=1}e^{i n \tilde{\varepsilon}}
 M(t_Q,t_{Q-n}) \right\}.
\end{align}
The second term in the third line in Eq.\eqref{Suppl:Current} can be disregarded because it is independent of the bias voltage and does not contribute to the tunneling conductance. The term $S(\varepsilon) + S(-\varepsilon)$ in Eq.~\eqref{Suppl:Current} is written in terms of the Majorana Green's function, and contains information of the non-commuting braiding operations. It yields 
\begin{align}
S(\varepsilon) + S(-\varepsilon)&= \sum^{[Q/4]}_{m=-[Q/4]} e^{i m \left(4 \tilde{\varepsilon} + \alpha\right)}. 
\end{align} 
The notation $[Q/4]$ denotes the integer part of the number $Q/4$ and we have used anti-commutation relations 
\begin{align}
\left\{\hat{\gamma}_1(t_q), \hat{\gamma}_1(t_{q'}) \right\} = 
\left\{
\begin{array}{cc}
2 e^{i m \alpha}  &  \text{for}~q-q'=4 m,\\
 0 & \text{otherwise}.
\end{array}
\right.\label{ACrelation}
\end{align}
The phase factor $e^{i m \alpha}$ with $\alpha = \pi$ comes from a $2\pi m$-rotation of the four MBSs, 
$\left(U^{\dagger}_c\right)^{4m} \gamma_j \left(U_c\right)^{4m} = (-1)^m \gamma_j$, and is physically 
due to crossing branch cuts emanating from the MBSs. In the limit $Q \rightarrow \infty$ 
(or $\tilde{t}-t_0 \rightarrow \infty$), we obtain
\begin{align}
\frac{d\bar{I}}{dV}= \frac{e^2}{h}\frac{\pi \hbar}{8 T_J k_B T} 
\sum_l T(\varepsilon_l)~\text{sech}^2\left(\frac{eV - \varepsilon_l}{2 k_B T}\right),
\end{align}
where $\varepsilon_l=\frac{\hbar}{4T_J}(2\pi l-\alpha)$ with integer $l$.  

\clearpage
\section*{B. Occupation number representation}\label{Numberbasis}

\begin{figure}
\centering
\includegraphics[width=0.7\columnwidth]{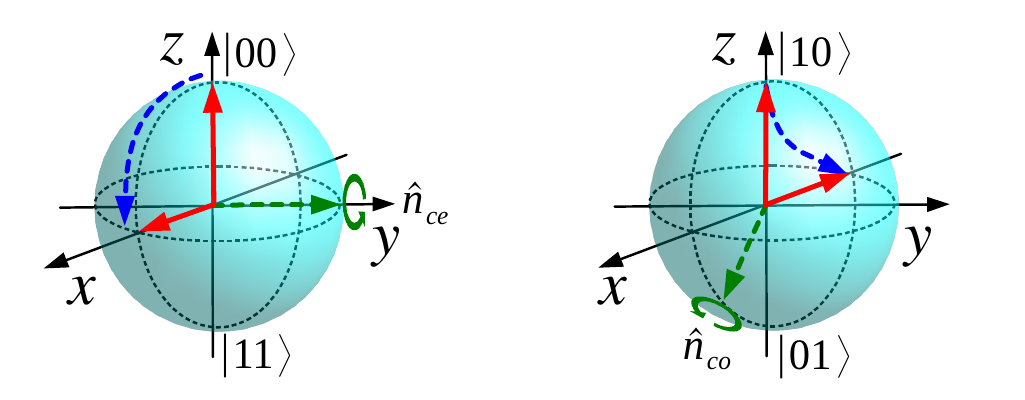}
\caption{Bloch sphere representation of Majorana rotation. The rotation of four MBSs by $\pi/2$ results in the rotation of Majorana qubit defined in the even parity space by $\pi/2$ about the direction $\hat{n}_{ce}$ (left sphere) and the rotation of Majorana qubit in the odd parity space by $\pi$ about the direction $\hat{n}_{co}$ (right sphere). The qubit rotations in the even and odd parity spaces which are intertwined by electron tunneling do not commute, leading to non-Abelian interference effect in the time-averaged tunneling conductance.}
\label{SFig2:Bloch}
\end{figure}

We describe the rotation of the four MBSs in occupation number space. We define two complex fermion operators, 
\begin{equation}
f_1 =\frac{\gamma_1 + i \gamma_2}{2}, \hspace{20pt} f_2 =\frac{\gamma_3 + i \gamma_4}{2},
\end{equation}
and four occupation number states which are degenerate at zero energy,
\begin{eqnarray}
\begin{split}
 &|00\rangle,  &&|10\rangle = f^{\dagger}_1|00\rangle, \\
 &|11\rangle = f^{\dagger}_1 f^{\dagger}_2 |00\rangle, &&|01\rangle = f^{\dagger}_2|00\rangle. 
\end{split}\label{Number_states}
\end{eqnarray}
Here, the state $|00\rangle$ is defined by $f_1|00\rangle =f_2|00\rangle =0$.
The two states in each fermion-occupation-number parity subspace form a qubit, $|00\rangle$ and $|11\rangle$ for the even and $|10\rangle$ and $ |01\rangle$ for the odd fermion parity subspace.
In the basis $\{|00\rangle, |11\rangle, |10\rangle, |01\rangle\}$, 
$U_c$ in the main text is represented as
\begin{align}
 U_c=
 \begin{pmatrix}
  U_{ce} & \bf{0} \\
  \bf{0} & U_{co}
 \end{pmatrix}
 =
 \begin{pmatrix}
  e^{-i\frac{\pi}{4} \hat{n}_{ce}\cdot \vec{\tau}} & \bf{0}  \\
  \bf{0}  & e^{-i \frac{\pi}{2} \hat{n}_{co}\cdot \vec{\tau}} 
 \end{pmatrix}, \label{C_Matrix}
\end{align}
where $U_{ce} (U_{co})$ is the evolution operator acting on the even (odd) parity space that rotates the qubit by $\pi/2 (\pi)$ about the direction of $\hat{n}_{ce} (\hat{n}_{co})$ 
given by 
\begin{align}
 \hat{n}_{ce}=(0,1,0), \hspace{10pt}
 \hat{n}_{co}=\frac{s}{\sqrt{2}}(-1,0,1). \label{UnitVector}
\end{align}
$\vec{\tau}=(\tau_x, \tau_y, \tau_z)$ are Pauli matrices acting on the qubit, 
and $\bf{0}$ is $2\times2$ null matrix. The qubit rotations induced by $U_{ce}$ and $U_{co}$ on the Bloch sphere are illustrated in Fig.~\ref{SFig2:Bloch}. 

The parity of the fermion occupation number is conserved in the transformation $U_c$.
$\bar{U}_c = \gamma_1 U_c \gamma_1$ in the same basis is represented by the interchange of $U_{ce}$ and $U_{co}$,
\begin{align}
 \bar{U}_c=\left(
 \begin{array}{cc}
  U_{co} & \bf{0} \\
  \bf{0} & U_{ce}
 \end{array}
\right).
\end{align}
We find that $U^{\dagger}_c$ and $\bar{U}_c$ (or $U_c$ and $\bar{U}_c$) do not commute,
\begin{align}
\left[\bar{U}_c, U^{\dagger}_c \right]
&=  
\begin{pmatrix}
[U_{co}, U^\dagger_{ce}]  & \bf{0} \\
\bf{0} &[U_{ce}, U^\dagger_{co}]
\end{pmatrix} \nonumber\\
&= s i 
\begin{pmatrix}
-\tau_x - \tau_z  & \bf{0} \\
\bf{0} & \tau_x + \tau_z
\end{pmatrix} \nonumber\\
&\neq 0. \label{Crelation}
\end{align}
This is indicative of the different braiding evolutions of world lines corresponding to the two operator products $\bar{U}_c U^{\dagger}_c$ and $U^{\dagger}_c  \bar{U}_c$.

\clearpage

\section*{C. Majorana wave functions}\label{waveftn}

We provide the details of the calculation of Majorana wave functions
$\Psi_{\text{M}j}({\mathbf r})$ with $j \in \{1,2,3,4 \}$ in a Corbino geometry topological Josephson junction. 
We solve the BdG equation $\mathcal{H}_{C} \Psi({\mathbf r}, \phi_1,\phi_2) 
= E \Psi({\mathbf r},\phi_1,\phi_2)$ for $E = 0$ and $\mu=0$. 
Hereafter, we use the dimensionless length scale $r$ normalized by $\xi = \hbar v_F / \Delta_0$. 
For $r<R$, the wave function $\Psi_{r<R}(r,\theta)$ is given by
\begin{align}
 \Psi_{r<R}(r,\theta) = \sum^{\infty}_{m=-\infty} a_m 
 \left(
 \begin{array}{cccc}
  e^{i m \theta} e^{i \phi_1/2} I_{m}(r)\\
  0\\
  0\\
  i e^{i (m+1) \theta} e^{-i \phi_1/2} I_{m+1}(r)
 \end{array}
 \right) +b_m
 \left(
 \begin{array}{cccc}
 0\\
 i e^{i (m+1) \theta} e^{i \phi_1/2} I_{m+1}(r)\\
 -e^{i m \theta} e^{-i \phi_1/2} I_{m}(r)\\
  0
 \end{array}
 \right),
\end{align}
where $I_m(r)$ is the modified Bessel function of the first kind, and $a_m$ and $b_m$ are coefficients. 
The wave function $\Psi_{r>R}(r,\theta)$ at $r>R$ is given by
\begin{align}
 \Psi_{r>R}(r,\theta) = \sum^{\infty}_{n=-\infty} c_n 
 \left(
 \begin{array}{cccc}
  i e^{i n \theta} e^{i \phi_2/2} r^{-2} K_{n+2}(r)\\
  0\\
  0\\
  e^{i (n+5) \theta} e^{-i \phi_2/2} r^{-2} K_{n+3}(r)
 \end{array}
 \right) +d_n
 \left(
 \begin{array}{cccc}
 0\\
 i e^{i (n+1) \theta} e^{i \phi_2/2} r^2 K_{n+3}(r)\\
 e^{i (n+4) \theta} e^{-i \phi_2/2} r^2 K_{n+2}(r)\\
  0
 \end{array}
 \right),
\end{align}
where $K_n(r)$ is the modified Bessel function of the second kind, and $c_n$ and $d_n$ are coefficients. 
We consider only the wave functions with spin down as those for spin up become non-normalizable solutions, and hence the coefficients $a_m$ and $c_n$ should be zero for all $m$ and $n$. 
In order to get the coefficients $b_m$ and $d_n$ we match the spin-down components at $r=R$, 
\begin{align}
\sum^{\infty}_{m=-\infty} b_m
 \left(
 \begin{array}{cccc}
 0\\
 i e^{i (m+1) \theta} e^{i \phi_1/2} I_{m+1}(R)\\
 -e^{i m \theta} e^{-i \phi_1/2} I_{m}(R)\\
  0
 \end{array}
 \right) =
 \sum^{\infty}_{n=-\infty} d_n
 \left(
 \begin{array}{cccc}
 0\\
 i e^{i (n+1) \theta} e^{i \phi_2/2} R^2 K_{n+3}(R)\\
 e^{i (n+4) \theta} e^{-i \phi_2/2} R^2 K_{n+2}(R)\\
  0
 \end{array}
 \right),
\end{align}
leading to 
\begin{align}
 b_l e^{i \phi_1/2} I_{l+1}(R) &= d_l e^{i \phi_2/2} R^2 K_{l+3}(R), \nonumber\\
 -b_{l+4} e^{-i \phi_1/2} I_{l+4}(R) &= d_l e^{-i \phi_2/2} R^2 K_{l+2}(R), \label{FtnMatching}
\end{align}
and the following recurrence relations,
\begin{align}
 b_{l+4} &= -e^{i(\phi_1 - \phi_2)} \frac{I_{l+1}(R) K_{l+2}(R)}{I_{l+4}(R) K_{l+3}(R)} b_{l}, \nonumber\\
 d_{l'+4} &= -e^{i(\phi_1 - \phi_2)} \frac{I_{l'+5}(R) K_{l'+2}(R)}{I_{l'+4}(R) K_{l'+7}(R)} d_{l'}, \label{RecurrentRelation}
\end{align}
where $l$ and $l'$ are integers.
From these recurrence relations we can construct four linearly independent solutions,
\begin{align}
 \Psi_{\eta}(r,\theta) &= \Theta(R-r) \sum^{\infty}_{m=-\infty} b_{4m+{\eta}} \Psi^{<}_{4m+{\eta}}(r,\theta) + \Theta(r-R) \sum^{\infty}_{n=-\infty} d_{4n+{\eta}} \Psi^{>}_{4n+{\eta}}(r,\theta) \nonumber\\
 &= b_{\eta} \Psi'_{\eta}(r,\theta), \label{app:sol1} 
 \end{align}
where ${\eta}\in \{-2,-1,0,1 \}$ and $\Psi'_{\eta}(r,\theta)$ are 
\begin{align}
 \Psi'_{\eta}(r,\theta) = \Theta(R-r) \sum^{\infty}_{m=-\infty} B_{m{\eta}} \Psi^{<}_{4m+{\eta}}(r,\theta) + \Theta(r-R) \sum^{\infty}_{n=-\infty} D_{n{\eta}} \Psi^{>}_{4n+{\eta}}(r,\theta).\label{app:sol2}
 \end{align}
Here the wave functions $\Psi^{<}_{4m+{\eta}}(r,\theta)$ at $r<R$ and $(\Psi^{>}_{4n+{\eta}}(r,\theta))$ at $r > R)$ are given by
\begin{align}
 \Psi^{<}_{4m+{\eta}}(r,\theta) &=
 \left(
 \begin{array}{cccc}
 0\\
 i e^{i (4m+{\eta}+1) \theta} e^{i \phi_1/2} I_{4m+{\eta}+1}(r)\\
 -e^{i (4m+{\eta}) \theta} e^{-i \phi_1/2} I_{4m+{\eta}}(r)\\
  0
 \end{array}
 \right), \nonumber\\
 \Psi^{>}_{4n+{\eta}}(r,\theta) &=
 \left(
 \begin{array}{cccc}
 0\\
 i e^{i (4n+{\eta}+1) \theta} e^{i \phi_2/2} r^2 K_{4n+{\eta}+3}(r)\\
 e^{i (4n+{\eta}+4) \theta} e^{-i \phi_2/2} r^2 K_{4n+{\eta}+2}(r)\\
  0
 \end{array}
 \right),
\end{align}
and coefficients $B_{m{\eta}}$ and $D_{n{\eta}}$ are  
\begin{align}
 B_{m{\eta}} &= 
 \left\{
 \begin{array}{ccc}
  (-1)^{m} e^{i m (\phi_1-\phi_2)} \left[\displaystyle \prod^{m}_{k=1} \frac{I_{4(k-1)+{\eta}+1}(R) K_{4(k-1)+{\eta}+2}(R)}{I_{4(k-1)+{\eta}+4}(R) K_{4(k-1)+{\eta}+3}(R)} \right] &&~\text{for}~ m \geq 1, \\ \\
  1  &&~\text{for}~ m = 0, \\ \\
  (-1)^{m} e^{i m (\phi_1-\phi_2)} \left[\displaystyle \prod^{-m}_{k=1} \frac{I_{-4(k-1)+{\eta}}(R) K_{-4(k-1)+{\eta}-1}(R)}{I_{-4(k-1)+{\eta}-3}(R) K_{-4(k-1)+{\eta}-2}(R)}\right] &&~\text{for}~ m \leq -1, \\
 \end{array}
 \right.  \nonumber\\ \nonumber\\
 D_{n{\eta}} &= 
 \left\{
 \begin{array}{ccc}
  (-1)^{n} e^{i (n+1/2) (\phi_1-\phi_2)} \frac{I_{{\eta}+1}(R)}{R^2 K_{{\eta}+3}(R)} \left[\displaystyle \prod^{n}_{k=1} \frac{I_{4(k-1)+{\eta}+5}(R) K_{4(k-1)+{\eta}+2}(R)}{I_{4(k-1)+{\eta}+4}(R) K_{4(k-1)+{\eta}+7}(R)} \right] &&~\text{for}~ n \geq 1, \\ \\
  e^{i (\phi_1-\phi_2)/2} \frac{I_{{\eta}+1}(R)}{R^2 K_{{\eta}+3}(R)}  &&~\text{for}~ n = 0, \\ \\
  (-1)^{n} e^{i (n+1/2) (\phi_1-\phi_2)} \frac{I_{{\eta}+1}(R)}{R^2 K_{{\eta}+3}(R)} \left[\displaystyle \prod^{-n}_{k=1} \frac{I_{-4(k-1)+{\eta}}(R) K_{-4(k-1)+{\eta}+3}(R)}{I_{-4(k-1)+{\eta}+1}(R) K_{-4(k-1)+{\eta}-2}(R)}\right] &&~\text{for}~ n \leq -1. \\
 \end{array}
 \right. \nonumber
\end{align}
\begin{figure}
\centering
\includegraphics[width=0.6\columnwidth]{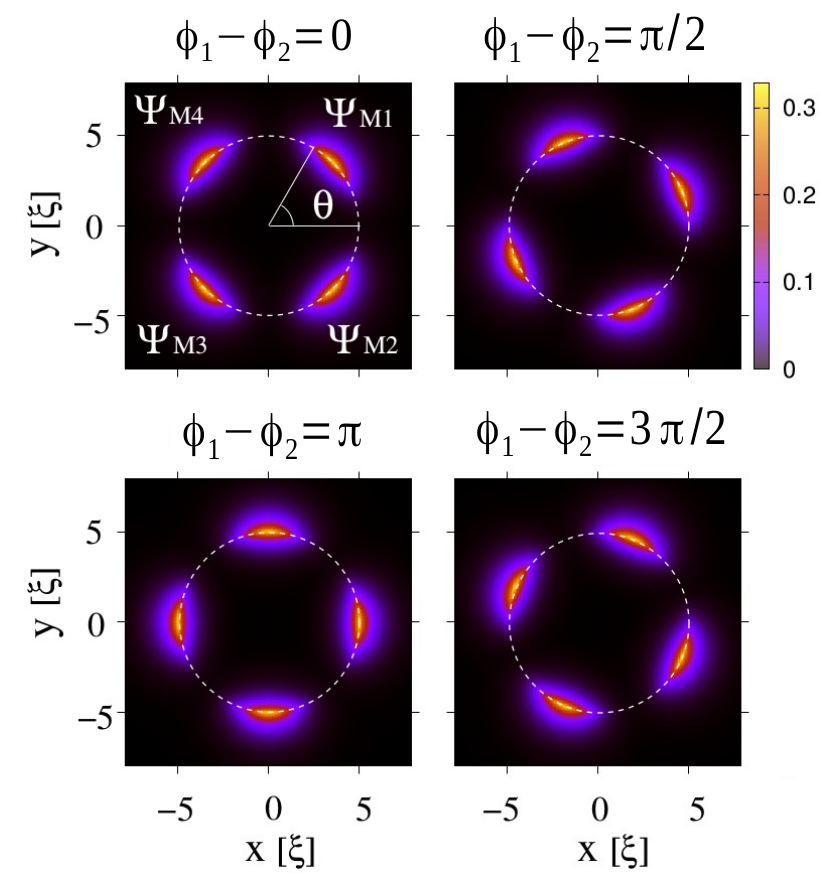}
\caption{Four Majorana wave functions. Probability densities of the four MBSs $|\Psi_{Mj}|^2$ in the Corbino geometry Josephson junction for different values of the superconducting phase difference $\phi_1-\phi_2$. The white dashed circle represents the interface between two superconductors, $S_1$ and $S_2$. The MBSs rotate by $\pi/2$ in a clockwise direction when $\phi_1-\phi_2$ varies by $2\pi$. The parameter used are the radius of the circle $R= 5\xi$ where $\xi=\hbar v_F/\Delta_0$ and the chemical potential $\mu=0$.}
\label{SFig1:MBS}
\end{figure}
By superposing the solutions $\Psi'_{\eta}(r,\theta)$ in Eq.~\eqref{app:sol2} and using particle-hole symmetry, 
we find four Majorana states $\Psi_{\text{M}j}$ satisfying $\Xi \Psi_{\text{M}j} = \Psi_{\text{M}j}$ where 
$\Xi = \sigma_y \tau_y \mathcal{C}$ is the particle-hole operator and $\mathcal{C}$ is
the operator for complex conjugation, see Fig.~\ref{SFig1:MBS}. They are given by 
\begin{align}
 \Psi_{\text{M}j}(r,\theta) =  \sum^{1}_{{\eta}=-2} \frac{1}{\sqrt{4 N_{\eta}}}\text{exp}\left[i \frac{\pi}{4} - i \left( {\eta}+\frac{1}{2}\right) \theta_j \right] \Psi'_{\eta}(r,\theta), \label{MBSs}
\end{align}
where the azimuthal angles $\theta = \theta_{j}$ at which $\Psi_{\text{M}j}$ 
are localized are given by $\theta_j = (3\pi - 2\pi j)/4 - (\phi_1-\phi_2)/4$ and $N_{\eta}$ are normalization constants such that 
\begin{align}
 N_{\eta} = \int d^2 r \Psi'^{\dagger}_{\eta}({\mathbf r}) \Psi'_{\eta}({\mathbf r}).
\end{align}

\clearpage

\section*{D. Floquet analysis}\label{Floquet} 

We confirm the conductance peak positions in the main text by using a Floquet analysis for four rotating MBSs including all orders in tunneling at zero temperature. 
This Floquet description is applicable as the time-dependent Hamiltonian discussed in the main text is periodic in time with periodicity $T_J$. See Ref.~\cite{app-Cayssol2013} for a short review on the Floquet formalism applied to topological insulators and Ref.~\cite{app-Park2015} in which two rotating MBSs are analyzed by using the Floquet formalism.

The Floquet Hamiltonian $H_F$ is defined by the time-independent Hamiltonian that 
would yield the same evolution as with $U_c$ in the main text after one period $T_J$,
\begin{align}
U_c &= e^{-\frac{i}{\hbar} H_F T_J}, \nonumber\\
H_F &= \frac{\hbar}{T_J} 
\left(
\begin{array}{cc}
  \frac{\pi}{4} \hat{n}_{ce}\cdot \vec{\tau} & \bf{0}  \\
  \bf{0}  &  \frac{\pi}{2} \hat{n}_{co}\cdot \vec{\tau} 
\end{array}\right) - \frac{2 \pi l \hbar}{T_J} \mathbb{I}, \label{Floquet_H1}
\end{align}
where $\hat{n}_{ce}$ and $\hat{n}_{co}$ are given in Eq.~\eqref{UnitVector}, and $l$ is an integer. 
The last term $-2\pi l \hbar/T_J \mathbb{I}$ in $H_F$ 
only shifts the energy levels and can be ignored for the moment. At the end of the calculation, we will 
restore this term. The representation of $H_F$ in terms of Majorana operators is  
\begin{align}
 H_F &= \frac{i}{\sqrt{2}} E_0 \bigg( s \gamma_1 \gamma_2 + \frac{1}{\sqrt{2}} \gamma_1 \gamma_3 -s \gamma_1 \gamma_4 + s \gamma_2 \gamma_3 -\frac{1}{\sqrt{2}} \gamma_2 \gamma_4 -s \gamma_3 \gamma_4\bigg) \nonumber\\
&= \frac{i}{2} \sum_{i\neq j} t_{ij} \gamma_i \gamma_j \label{Floquet_H2}
\end{align}
where $E_0 = \pi \hbar/(4 T_J)$ and the subindices $i$ and $j$ range from $1$ to $4$.
$t_{ij}$ describing the effective coupling between MBSs caused by the rotation is 
the $(i,j)$ component of the matrix ${\bf t}$ in the basis $(\gamma_1, \gamma_2, \gamma_3, \gamma_4)$, 
\begin{align}
 {\bf t} &= \frac{E_0}{\sqrt{2}} 
 \begin{pmatrix}
  0 & s & \frac{1}{\sqrt{2}} & -s \\
  -s & 0 & s & -\frac{1}{\sqrt{2}} \\
  -\frac{1}{\sqrt{2}} & -s & 0 & -s \\
  s & \frac{1}{\sqrt{2}} & s & 0 
 \end{pmatrix}.
\end{align} 

We calculate the differential conductance of the metal tip coupled to the Majorana network 
described by $H_F$ following the Keldysh technique calculation used in Ref.~\cite{app-Flensberg2010}. 
We consider the case where the tip is coupled only to $\gamma_1$. 
The differential conductance then is given by the formula 
\begin{align}
 \frac{dI}{dV} = \frac{2 e^2}{h} \int d\omega \Gamma \text{Im}[G^R_{11}(\omega)]\frac{d}{d\omega} n_F(\omega-eV), \label{Floquet_didv}
\end{align}
where $G^R_{11}(\omega)$ is the $(1,1)$ component of 
the $4 \times 4$ matrix ${\bf G}^R(\omega)$ given by  
\begin{align}
 {\bf G}^R(\omega)=2[\omega - 2i {\bf t} +2i {\bf \Gamma}]^{-1},
\end{align}
where ${\bf \Gamma}$ describes the tip-MBS tunneling which is the $4 \times 4$ matrix whose components are given by $({\bf \Gamma})_{ij} = \Gamma \delta_{1i}\delta_{1j}$.
$G^R_{11}(\omega)$ is then computed as
\begin{align}
 G^{R}_{11}(\omega) &= \frac{2 \omega (\omega^2 - 5 E^2_0)}{(\omega^2 - E^2_0)(\omega^2 - 9 E^2_0) + 2 i \omega \Gamma (\omega^2 - 5 E^2_0)}.
\end{align}
Substituting this into Eq.~\eqref{Floquet_didv} gives the differential conductance at zero temperature, 
\begin{align}
 \frac{dI}{dV} =\frac{2 e^2}{h} \left[1+ \frac{((eV)^2 - E^2_0)^2 ((eV)^2 - 9 E^2_0)^2}{4 (eV)^2 \Gamma^2 ((eV)^2 - 5 E^2_0)^2} \right]^{-1}
\end{align}
which exhibits peaks at $eV = \pm E_0$ and $\pm 3 E_0$. 
If we restore the term $-2\pi l \hbar/T_J$ in Eq.~\eqref{Floquet_H1}, the peaks would be at 
\begin{align}
 eV = \pm \frac{\pi \hbar}{4 T_J} -\frac{2 \pi l \hbar}{T_J},~\pm \frac{3 \pi \hbar}{4 T_J} -\frac{2 \pi l \hbar}{T_J}.
\end{align}
Therefore, we conclude that the Floquet theory gives a consistent result with 
the time-averaged differential conductance shown in Fig. 3 in the main text.

\clearpage

\section*{E. Case of commuting braiding operations}

For an unambiguous demonstration of the relation of the conductance peak positions to the presence of non-Abelian operations, we consider similar tunneling experiments, that is, a tip is coupled to $\gamma_1$ at $t=t_0$ and the system Hamiltonian is periodic in time $T_J$, but with commutating operations of four MBSs. We explicitly show that the resulting conductance peak positions are different from those in Eq.~(12) in the main text. 

\begin{figure}
\centering
\includegraphics[width=0.6\columnwidth]{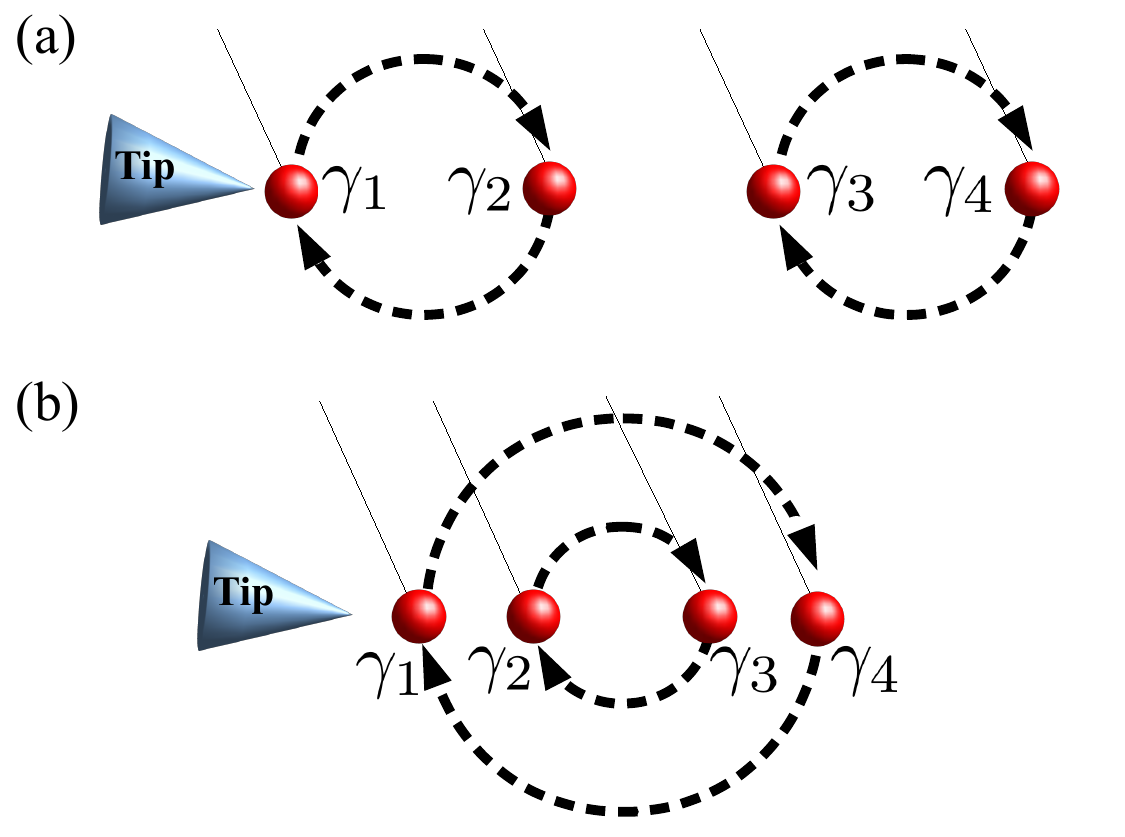}
\caption{Tunneling experiments with commuting braiding operators. Schematic illustration of possible examples of 
tunneling experiments into MBS resulting in commuting braiding operators $W$ and $\bar{W}$, see Eq.~\eqref{W_operator}. 
(a) Two pairs of MBSs are rotating in different circles being well-separated 
from each other with tunneling from a tip to $\gamma_1$ at every half rotation. 
Transformation from $(\gamma_{1}, \gamma_{2}, \gamma_{3}, \gamma_{4})$ to 
$(-\gamma_{2}, \gamma_{1}, -\gamma_{4}, \gamma_{3})$ by a half-rotation are 
described by $\hat{n}_{w} = \hat{z}, \beta = \pi/2$, and $\beta'=0$ of $W$.   
(b) Similar to the case of (a), but with two pairs of MBSs in concentric circles 
with different radii. They transform from $(\gamma_{1}, \gamma_{2}, \gamma_{3}, \gamma_{4})$ 
to $(-\gamma_{4}, \gamma_{3}, -\gamma_{2}, \gamma_{1})$ by a half-rotation, and 
corresponding parameter values of $W$ are given by $\hat{n}_{w} = \hat{x}, \beta = 0$ and $\beta'=-\pi/2$. 
Thin lines in (a) and (b) denote branch cuts.}
\label{SFig3:Commute}
\end{figure}

Let us introduce two different evolution operators, $W$ and $\bar{W}=\gamma_1 W \gamma_1$, similar to the parity-conserving braiding operator $U_c$ and the tunneling-assisted braiding operator $\bar{U}_c$, respectively. The only difference compared to $U_c$ and $\bar{U}_c$ is that 
$W$ and $\bar{W}$ commute such that $[W, \bar{W}] = 0$. This commutativity condition 
allows us to find the generic form of $W$ (and thus of $\bar{W}$) to be,  
\begin{align}
W &= 
\begin{pmatrix}
W_e & \bf{0} \\
\bf{0} & W_o
\end{pmatrix} = 
\begin{pmatrix}
e^{i \beta \hat{n}_{w} \cdot \vec{\tau}} & \bf{0} \\
\bf{0} & e^{i \beta' \hat{n}_{w} \cdot \vec{\tau}}
\end{pmatrix}, \label{W_operator}
\end{align}  
up to an overall phase factor which does not affect the tunneling current. The rigorous derivation of this form of $W$ is given in the next section.
$\bar{W}$ is obtained by interchanging $W_e$ and $W_o$ in the $W$ matrix. Here the general commuting braiding operations are characterized by the unit vector $\hat{n}_{w}$ and angles $\beta$ and $\beta'$. Two related situations are drawn in Fig.~\ref{SFig3:Commute}. 
 
The interfering terms in this case commute and thus can be expressed as 
\begin{align}
\sum^{\infty}_{n=1} e^{i n \tilde{\varepsilon}} 
W_{o}^{n} \left(W^{\dagger}_{e}\right)^{n} =  
\sum^{\infty}_{n=1} e^{i n \tilde{\varepsilon}} e^{i n (\beta'-\beta) \hat{n}_{w} \cdot \vec{\tau}}, \label{Basis-Trans2}
\end{align}
indicating that the relative dynamics between $n$ and $n+1$ cycles adds a phase factor to the eigenstate of $\hat{n}_{w} \cdot \vec{\tau}$. In order to find its consequence, we calculate the anti-commutation relation,  
\begin{align}
\left\{\hat{\gamma}_1(t_q), \hat{\gamma}_1(t_{q'}) \right\} &=
\left(\bar{W} W^{\dagger} \right)^{(q-q')} + \left(W \bar{W}^{\dagger} \right)^{(q-q')} \nonumber\\
&= 2~ \text{cos}\left[(q-q') (\beta -\beta') \right], \label{ACrelation_2}
\end{align}
where $\hat{\gamma}_1(t_q) = \left(W^{\dagger}\right)^{q} \gamma_1 W^{q}$. 
By substituting this into $S(\varepsilon)+S(-\varepsilon)$ of Eq.~\eqref{Suppl:Current},
the peak positions of the conductance in the low-bias voltage regime are found as   
\begin{align}
eV = \frac{\hbar}{T_J}\left[2\pi j \pm \left( \beta - \beta'\right) \right],
\end{align} 
where $j$ is an integer. If $0 \leq |\beta-\beta'| < \pi$, the peak separations 
$2 |\beta-\beta'| \hbar/T_J$ and $(2 \pi - 2 |\beta-\beta'|) \hbar/T_J$ appear alternately. 
If $\pi \leq |\beta-\beta'| < 2\pi$, then the separations of $(2 |\beta-\beta'| - 2 \pi) \hbar/T_J$ 
and $(4 \pi - 2 |\beta-\beta'|) \hbar/T_J$ are seen alternately. 
Note that for any value of $|\beta-\beta'|$, these peak configurations cannot give rise to 
the results shown in Eq. (12) in the main text, manifesting the noncommutative structure of non-Abelian statistics.  

\section*{F. Derivation of a generic form of $W$} \label{Wproof}

We argue that the generic form of the matrix $W$  
satisfying $\left[W, \bar{W} \right]=0$, shown in Eq.~\eqref{W_operator}, is 
\begin{align}
W= 
\begin{pmatrix}
W_e & \bf{0} \\
\bf{0} & W_o
\end{pmatrix}= 
\begin{pmatrix}
e^{i \beta \hat{n}_w \cdot \vec{\tau}} & \bf{0} \\
\bf{0} & e^{i \beta' \hat{n}_w \cdot \vec{\tau}}
\end{pmatrix} \label{W-matrix}
\end{align} 
where 
\begin{align}
\bar{W} = \gamma_1(t_0) W \gamma_1(t_0) = 
\begin{pmatrix}
{\bf 0} & \sigma_0 \\
\sigma_0 & {\bf 0}
\end{pmatrix}
\begin{pmatrix}
W_e & {\bf 0} \\
{\bf 0} & W_o
\end{pmatrix}
\begin{pmatrix}
{\bf 0} & \sigma_0 \\
\sigma_0 & {\bf 0}
\end{pmatrix}=
\begin{pmatrix}
W_o & {\bf 0} \\
{\bf 0} & W_e
\end{pmatrix}.
\end{align} 
Here $\sigma_0$ is the $2 \times 2$ identity matrix and ${\bf 0}$ is the null matrix. 
In this appendix, we prove this argument. 
Let us define $W$ as 
\begin{align}
W= 
\begin{pmatrix}
W_e & \bf{0} \\
\bf{0} & W_o
\end{pmatrix}= 
\begin{pmatrix}
e^{i \beta \hat{n}_e \cdot \vec{\tau}} & \bf{0} \\
\bf{0} & e^{i \beta' \hat{n}_o \cdot \vec{\tau}}
\end{pmatrix}=
\begin{pmatrix}
\text{cos} \beta + i~ \hat{n}_e \cdot \vec{\tau}~ \text{sin} \beta & \bf{0} \\
\bf{0} & \text{cos} \beta' + i~ \hat{n}_o \cdot \vec{\tau}~ \text{sin} \beta'
\end{pmatrix},
\end{align}
where the unit vectors $\hat{n}_e$ and $\hat{n}_o$ are 
\begin{align}
\hat{n}_e = (e_x, e_y, e_z), \hspace{20pt}
\hat{n}_o = (o_x, o_y, o_z).
\end{align}
$\bar{W}$ is then given by 
\begin{align}
\bar{W}= 
\begin{pmatrix}
W_o & \bf{0} \\
\bf{0} & W_e
\end{pmatrix}= 
\begin{pmatrix}
e^{i \beta' \hat{n}_o \cdot \vec{\tau}} & \bf{0} \\
\bf{0} & e^{i \beta \hat{n}_e \cdot \vec{\tau}}
\end{pmatrix}.
\end{align}

{\bf Case 1.} $\beta$ or $\beta'$ is equal to a multiple of $\pi$. 

We show that the form of $W$ in Eq.~\eqref{W-matrix} holds for the following cases 
\begin{itemize} 
\item $\beta = m \pi$ and $\beta' \neq n \pi$ 
\item $\beta \neq m \pi$ and $\beta' = n \pi$ 
\item $\beta = m \pi$ and $\beta' = n \pi$,
\end{itemize}
where $m$ and $n$ are integers. 
It is enough to consider the first case where $\beta = m \pi$ and $\beta' \neq n \pi$ 
as the proofs for the other two cases are similar. 
The matrix $W_e$ in this case is 
\begin{align}
W_e = \text{cos} \beta + i~ \hat{n}_e \cdot \vec{\tau}~ \text{sin} \beta = (-1)^{m} \sigma_0,
\end{align}
independent of $\hat{n}_e$ and the commutation relation between $W$ and $\bar{W}$ is zero,
\begin{align}
\left[ W, \bar{W} \right] = 
\begin{pmatrix}
\left[W_e, W_o \right] & \bf{0} \\
\bf{0} & \left[W_o, W_e \right]
\end{pmatrix} 
=0.
\end{align} 
We can rewrite $W_e$ as 
\begin{align}
W_e = \text{cos} \beta + i~ \hat{n}_o \cdot \vec{\tau}~ \text{sin} \beta = e^{i \beta \hat{n}_o \cdot \vec{\tau}},
\end{align}
which is a valid expression if $\beta = m \pi$. 
Therefore, $W$ is written as 
\begin{align}
W= 
\begin{pmatrix}
W_e & \bf{0} \\
\bf{0} & W_o
\end{pmatrix}= 
\begin{pmatrix}
e^{i \beta \hat{n}_o \cdot \vec{\tau}} & \bf{0} \\
\bf{0} & e^{i \beta' \hat{n}_o \cdot \vec{\tau}}
\end{pmatrix},
\end{align}
which completes the proof in this case by changing the notation $\hat{n}_o$ by $\hat{n}_w$. 

{\bf Case 2.} $\beta \neq m \pi$ and $\beta' \neq n \pi$. 
 
In this case, we solve the problem 
\begin{align}
\left[ W, \bar{W} \right] = 
\begin{pmatrix}
\left[W_e, W_o \right] & \bf{0} \\
\bf{0} & \left[W_o, W_e \right]
\end{pmatrix} 
=0.
\end{align}
Specifically, we need to solve 
\begin{align}
\left[ W, \bar{W} \right] &= \left[\text{cos} \beta + i~ \hat{n}_e \cdot \vec{\tau}~ \text{sin} \beta, 
\text{cos} \beta' + i~ \hat{n}_o \cdot \vec{\tau}~ \text{sin} \beta' \right]  \nonumber\\
&= - \text{sin} \beta \text{sin} \beta' \left[\hat{n}_e \cdot \vec{\tau}, \hat{n}_0 \cdot \vec{\tau}  \right] \nonumber\\
&=0.
\end{align}
Because $\text{sin} \beta \text{sin} \beta' \neq 0$ in this case, $ \left[\hat{n}_e \cdot \vec{\tau}, \hat{n}_0 \cdot \vec{\tau}  \right]$ 
should be zero, which leads to 
\begin{align}
 \left[\hat{n}_e \cdot \vec{\tau}, \hat{n}_0 \cdot \vec{\tau}  \right] 
 &= 2 i \left[ (e_y o_z - e_z o_y) \tau_x + (e_z o_x - e_x o_z) \tau_y + (e_x o_y - e_y o_x) \tau_x\right] \nonumber\\
 &= 0.  
\end{align}
As the Pauli matrices $\tau_{x,y,z}$ form an orthogonal basis, we have 
\begin{align}
e_y o_z - e_z o_y &= 0, \nonumber\\
e_z o_x - e_x o_z &= 0, \nonumber\\
e_x o_y - e_y o_x &= 0,
\end{align}
and 
\begin{align}
\frac{o_x}{e_x} = \frac{o_y}{e_y} = \frac{o_z}{e_z},
\end{align}
which allows us to find
\begin{align}
(o_x, o_y, o_z) = s (e_x, e_y, e_z), 
\end{align}
where $s = 1$ or $-1$. 
Therefore, $\hat{n}_o = s \hat{n_e}$ and $W$ in this case is obtained by 
\begin{align}
W=  
\begin{pmatrix}
e^{i \beta \hat{n}_e \cdot \vec{\tau}} & \bf{0} \\
\bf{0} & e^{i \beta' \hat{n}_o \cdot \vec{\tau}}
\end{pmatrix}
=\begin{pmatrix}
e^{i \beta \hat{n}_e \cdot \vec{\tau}} & \bf{0} \\
\bf{0} & e^{i \beta' s \hat{n}_e \cdot \vec{\tau}}
\end{pmatrix}. 
\end{align}
By redefining notations $s \beta'$ and $\hat{n}_e$ by $\beta'$ and $\hat{n}_w$, 
we obtain Eq.~\eqref{W-matrix}.


\begin{thebibliography}{50}
\bibitem{app1} In our model Hamiltonian, we assume that the distance between $S_1$ and $S_2$ is zero, and the radius $R$ of the inner superconductor satisfies $\xi \ll R \ll \Lambda$ so that the vector potential of a magnetic field can be neglected, where $\xi$ is the superconducting coherence length and $\Lambda$ is the Pearl length. We also assume a short circumference of the junction, $2 \pi R \ll \lambda_J$ with the Josephson penetration depth $\lambda_J$ (see Refs.~\cite{Park2015,Clem2010}).
\end{thebibliography}
\end{document}